\documentclass[twocolumn,amsmath,floatfix,prb,aps]{revtex4-1}
\usepackage{color}
\usepackage{amsmath}
\usepackage{pifont}   
\usepackage{graphicx} 
\usepackage{dcolumn}  
\usepackage{bm}       
\usepackage{amsfonts} 
\usepackage{amssymb}  
\usepackage{multirow} 
\usepackage{siunitx}

\usepackage[english]{babel}
\usepackage[autostyle, english = american]{csquotes}
\MakeOuterQuote{"}

\usepackage{placeins}

\usepackage{braket} 

\usepackage{bbm}

    \renewcommand{\v}[1]{\bm{\mathrm{#1}}}
    \newcommand{\m}[1]{\bm{\mathsf{#1}}}

\newcommand{\bk}{{\bf k}}

\begin{document}

\title{Mapping the energy-time landscape of spins with helical X-rays}

\author{N. Pontius$^{1}$}
\author{J.~K. Dewhurst$^2$}
\author{C. Sch\"ussler-Langeheine$^{1}$}
\author{S. Jana$^{3}$}
\author{C. v. Korff Schmising$^3$}
\author{S. Eisebitt$^3$}
\author{S. Shallcross$^3$}
\author{S. Sharma$^3$}
\email{sharma@mbi-berlin.de}
\affiliation{1 Helmholtz-Zentrum Berlin f{\"u}r Materialien und Energie GmbH, Albert-Einstein-Stra{\ss}e 15, 12489 Berlin, Germany}
\affiliation{2 Max-Planck-Institut fur Mikrostrukturphysik Weinberg 2, D-06120 Halle, Germany}
\affiliation{3 Max-Born-Institute for Non-linear Optics and Short Pulse Spectroscopy, Max-Born Strasse 2A, 12489 Berlin, Germany}

\date{\today}

\begin{abstract}
Unveiling the key mechanisms that determine optically driven spin dynamics is essential both to probe the fundamental nature of ultrafast light-matter interactions, but also to drive future technologies of smaller, faster, and more energy efficient devices. Essential to this task is the ability to use experimental spectroscopic tools to evidence the underlying energy- and spin-resolved dynamics of non-equilibrium electron occupations.  In this joint theory and experimental work, we demonstrate that ultrafast helicity-dependent soft X-ray absorption spectroscopy (HXAS) allows access to spin-, time- and energy specific state occupation after optical excitation.  We apply this method to the prototype transition metal ferromagnet cobalt and find convincing agreement between theory and experiment. The richly structured energy-resolved spin dynamics unveil the subtle interplay and characteristic time scales of optical excitation and spin-orbit induced spin-flip transitions in this material: the spin moment integrated in an energy window below the Fermi level first exhibits an ultrafast increase as minority carriers are excited by the laser pulse, before it is reduced as spin-flip process in highly localized, low energy states start to dominate. The results of this study demonstrate the power of element specific transient HXAS, placing it as a potential new tool for identifying and determining the role of fundamental processes in optically driven spin dynamics in magnetic materials.
\end{abstract}

\maketitle

\section{Introduction}

The macroscopic spin and charge state of solids can be manipulated by light pulses on few femtosecond \cite{Bigot2009, Siegrist2019} to picosecond time scales \cite{Kirilyuk2010,Carva2017, Kimel2020, Jeppson2021}. A wealth of fundamental new phenomena have been uncovered in this "femtoscape", including  all-optical magnetization reversal \cite{Stanciu2007,Radu2011} and laser-driven spin transport \cite{Battiato2010,Eschenlohr2013,dewhurst2018}, effects that carry profound implications for technological application. Progress in the fields of femtomagnetism, ultrafast valleytronics\cite{Jin2018} and ultrafast spintronics, ultimately rely on the ability of experiments to decode the complex and non-equilibrium population of spin and charge on ultrafast time scales. Particularly important for building a microscopic understanding of such processes is the ability to obtain, from spectroscopy experiments, a state and element-selective view of the spin and charge dynamics, i.e., at what energies, and from which atoms, is spin optically excited to and from during the time evolving magnetic state.

Such charge and spin excitations are determined by the pump pulse parameters and the nature of the magnetic material, as encoded in the spin-resolved density of states (DOS). In a magnetic material possessing nearly full majority bands the dominant available optical excitations are from minority states below to minority states above the Fermi level ($E_F$), a process that leaves the macroscopic magnetic state unchanged. Concurrently occurring spin-flip transitions will, however, reoccupy these minority states leading to the observed ultrafast demagnetization\cite{krieger2015,ZH00}. In alloys and hetero-structures differences between the local DOS of distinct elemental species may themselves result in the laser induced transfer of spin from one atomic site to the other, adding a further layer of complexity to the spin dynamics\cite{dewhurst2018,Siegrist2019,hofherr2020, Willems2020a, Tengdin2020,Golias2021}. Assessing the role these basic principles of optical excitation and spin flips play in the complex landscape of ultrafast demagnetization is of crucial importance for understanding, and possibly controlling, the remarkable physics of ultrafast magnetization dynamics.

To unravel the complex pathways of non-equilibrium charge within a site-, energy- and spin dependent landscape during ultrafast demagnetization, requires techniques capable of resolving these degrees of freedom. Some of this information can be extracted from spin and time dependent angle-resolved photoemission spectroscopy \cite{Rhie2003,Gort2018, Eich2017a, Frietsch2020}, which essentially measures the transient electronic spectrum. However, such experiments are not in general element specific and, in addition, are strongest when probing bands and momentum directions on surface/interface and around the Fermi energy, not when providing a global picture (i.e. fully energy resolved spectrum of the entire bulk of the solid) of spin and charge excitation. Magnetic circular dichroism (MCD) experiments, on the other hand, provide an element specific tool that readily offers access to transient magnetization \cite{Stamm2007a, Boeglin2010} separated into the contribution from spin and orbital moments \cite{Thole1992,Carra1993,Stohr2006,Stamm2010, Bergeard2014, Hennecke2019a}. 
Such experiments typically employ the XMCD contrast at fixed photon energy to probe magnetization dynamics, with recently a few experiments investigating spectral changes in absorption as a possible route to probe energy dependent magnetization dynamics \cite{Higley2019,Willems2020a,Hennes2020a,Rosner2020,LeGuyader2022}. While intriguing and as yet unexplained effects have been observed, a direct link to transient energy and spin-dependent electron occupations has not been established. 

In this joint theory and experimental work, we demonstrate that transient helicity-dependent X-ray absorption spectroscopy (HXAS), the "raw data" of XMCD experiments that is typically discarded, encodes detailed information of the energy- spin- and species-resolved electronic transitions that occur both during the optical excitation as well as in the subsequent relaxation processes. In order to extract this information we provide a method based on well founded assumptions of the underlying optical transitions, requiring for its facilitation the {\it ab-initio} determination of 4 parameters per material system.
Employing this tool and with an example of Co, we unveil the energy and spin resolved dynamics of laser pumped demagnetization, revealing the pathways in energy-spin space in which the ultrafast loss of moment occurs. We find a richly structured energy resolved spin dynamics with, remarkably, the spin moment below the Fermi level first exhibiting an increase after the laser pulse (due to excitation of minority charge), before falling as spin-flip process in highly localized low energy states dominate. In contrast, the magnetic moment above Fermi energy shows a continuous increase in magnitude (but with the moment pointing in the opposite direction) from the outset of the laser pulse.

\section{HXAS, XMCD and going beyond}

\begin{figure}[ht]
\includegraphics[width=\columnwidth, clip]{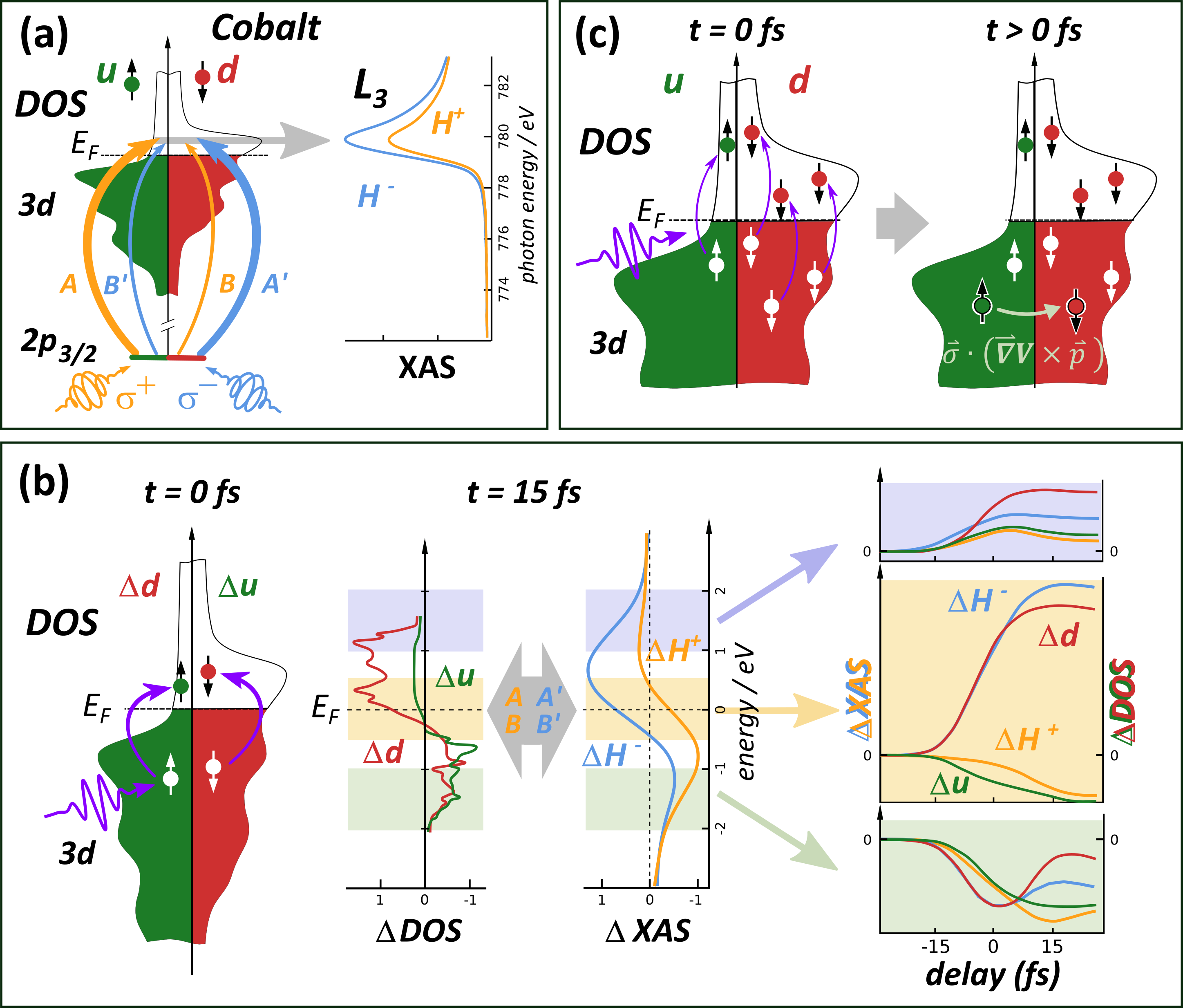}
\caption{
(a) Schematic illustration of helicity dependent X-ray absorption (HXAS) for the L$_{3}$ absorption resonance in Cobalt in which electrons are promoted from the $2p_{3/2}$ core level into the spin polarized $3d$ unoccupied valence levels. Light of positive helicity yields the yellow ($H^+$) XAS, negative helicity the blue ($X^-$) XAS. (b) Helicity dependent transient X-ray absorption. Laser induced changes to the spin dependent state occupation $\Delta$DOS, shown for majority ($\Delta u$) and minority ($\Delta d$) channels, are reflected in changes to the helicity dependent X-ray absorption $\Delta$XAS, with these two connected through the spin dependent excitation probabilities $A$, $B$, $A’$, and $B’$. Integration over the coloured energy regions yields the time, energy, and spin resolved changes in state occupation and XAS, shown on the right hand side. (c) Illustration of the spin-orbit spin flip process in ferromagnetic Cobalt  after optical excitation. The stronger optical laser excitation of the minority electrons creates localized holes below $E_F$, enhancing subsequent spin-orbit coupling induced transitions, in turn leading to an increased spin flip rate and demagnetization at early times.
}\label{fig:scheme}
\end{figure}

When circular polarized X-ray or XUV photons are transmitted through a magnetic material the absorption strength of excitations from spin-orbit split core levels to the valence bands depends on the sign of the helicity, i.e. the material shows circular dichroism. A specific example of the absorption of circular polarized X-rays is the $L_{3}$ resonance (excitations from the $2p_{3/2}$ core level to the 3$d$ valence band) in ferromagnetic Cobalt, illustrated in Fig. \ref{fig:scheme}a. This response function represents a measure of magnetic moment of the material: light of positive helicity primarily couples to spin-up (majority spin) electrons (leading to a response function we call $H^+(E)$), while light of negative helicity couples predominantly to spin-down (minority spin) electrons (leading to a response function we call $H^-(E)$)\cite{Stohr2006}. The absorption probability for a particular spin channel depends additionally on the available unoccupied valence states of the respective spin. This allows for "counting" of available valence states of specific spin, which is a direct measure of the spin polarization in the case of isolated atoms. To obtain this spin polarization from $H^+(E)$ and $H^-(E)$ for a solid one uses the normalized difference between both response functions, which yields the X-ray circular dichroism (XMCD) spectrum, together with the XMCD sum-rules (i.e. integration of this spectrum over the $L_{3}$ and the $L_{2}$ resonances in case of $3d$ elements like Co)\cite{Thole1992, Carra1993,chen1995,Stohr2006}. 
Element specificity of the XMCD response comes from the fact that the energies of these core levels strongly vary with the atomic species and can therefore be energetically separated.  

In the transient case occupations of bands above and below the Fermi energy evolve as charge is excited by the laser pump pulse (compare Fig.~\ref{fig:scheme}b). Subsequent dynamical processes such as spin-orbit scattering, reduce the local spin moments and so further change these occupations. The XMCD spectra in combination with the sum rules\cite{chen1995} can then be employed to obtain the resulting dynamics of the magnetic moment. This procedure has been enormously successful, and has been used to map early time spin dynamics in a number of magnets including species resolved magnetic order in multi-component magnets\cite{Stamm2010,Radu2011,Bergeard2014, Hennecke2019a}. However, what is not readily accessible, yet is of great importance, is information concerning energy and time dependent changes of a specific spin channel (and not just the integrated difference as represented by the magnetic moment). 

In obtaining magnetization dynamics from the XMCD spectra, the individual response functions to helical light, $H^+$ and $H^-$, are discarded. As we will now show, hidden within $H^+$ and $H^-$ lies detailed information on spin and energy resolved changes in the electronic structure of dynamical magnetic systems.

The helicity dependent absorption can be written as:

\begin{eqnarray}
H^{\pm}(E,t) &=& H_{\uparrow}^{\pm}(E,t) + H_{\downarrow}^{\pm}(E,t) \nonumber\\
&=&\Im \sum_{i,f} \Bigg[ p^{\pm}_{i,f,\uparrow}(t)\frac{(n_{i,\uparrow}-n_{f,\uparrow})(t)}{\Delta\epsilon_{if,\uparrow}(t)-E+i\eta} \nonumber \\
&+& p^{\pm}_{i,f,\downarrow}(t) \frac{(n_{i,\downarrow}-n_{f,\downarrow})(t)}{\Delta\epsilon_{if,\downarrow}(t)-E+i\eta}\Bigg] 
\label{hpmc}
\end{eqnarray}
where in the first line we have decomposed the helicity dependent absorption into the spin projected responses $H^{\pm}_\sigma(E,t)$, and in the second line give the corresponding linear response formulae. In these latter equations $i$ and $f$ are joint indices of state and {\bf k}-point and represent the initial and the final state respectively, $\uparrow$ and $\downarrow$ refer to the majority and minority spins, and $p^{\pm}_{if}$ are the transition probabilities for transitions from state $i$ to $f$ with left/right circularly polarized light. $\epsilon_i$ is the energy and $n_i$ the occupation of the $i$'th state. 
In the simplifying limit of a two state model, Eq.~\ref{hpmc} can be written as

\begin{eqnarray}
a(t)u(E,t)+b(t)d(E,t)&=H^+(E,t) \label{hxas0} \\
a'(t)u(E,t)+b'(t)d(E,t)&=H^-(E,t), \nonumber
\end{eqnarray}
i.e. in a way that separates the transition element information from that of state occupation and DOS (with $a(t) = p^+_{i f\uparrow}$, $b(t) =  p^+_{i f\downarrow}$, $a'(t) = p^-_{i f\uparrow}$, $b'(t) =  p^-_{i f\downarrow}$, and $u(E,t) = \Im (n_{i_\uparrow}-n_{f\uparrow})/(\Delta\epsilon_{if,\uparrow}-E+i\eta))$, and $d(E,t) = \Im (n_{i_\downarrow}-n_{f\downarrow})/(\Delta\epsilon_{if,\downarrow}-E+i\eta))$. The latter two quantities, $u(E,d)$ and $d(E,d)$, represent the DOS and occupation from which transient spin resolved charge can be extracted.

\begin{figure}[ht]
\includegraphics[width=\columnwidth, clip]{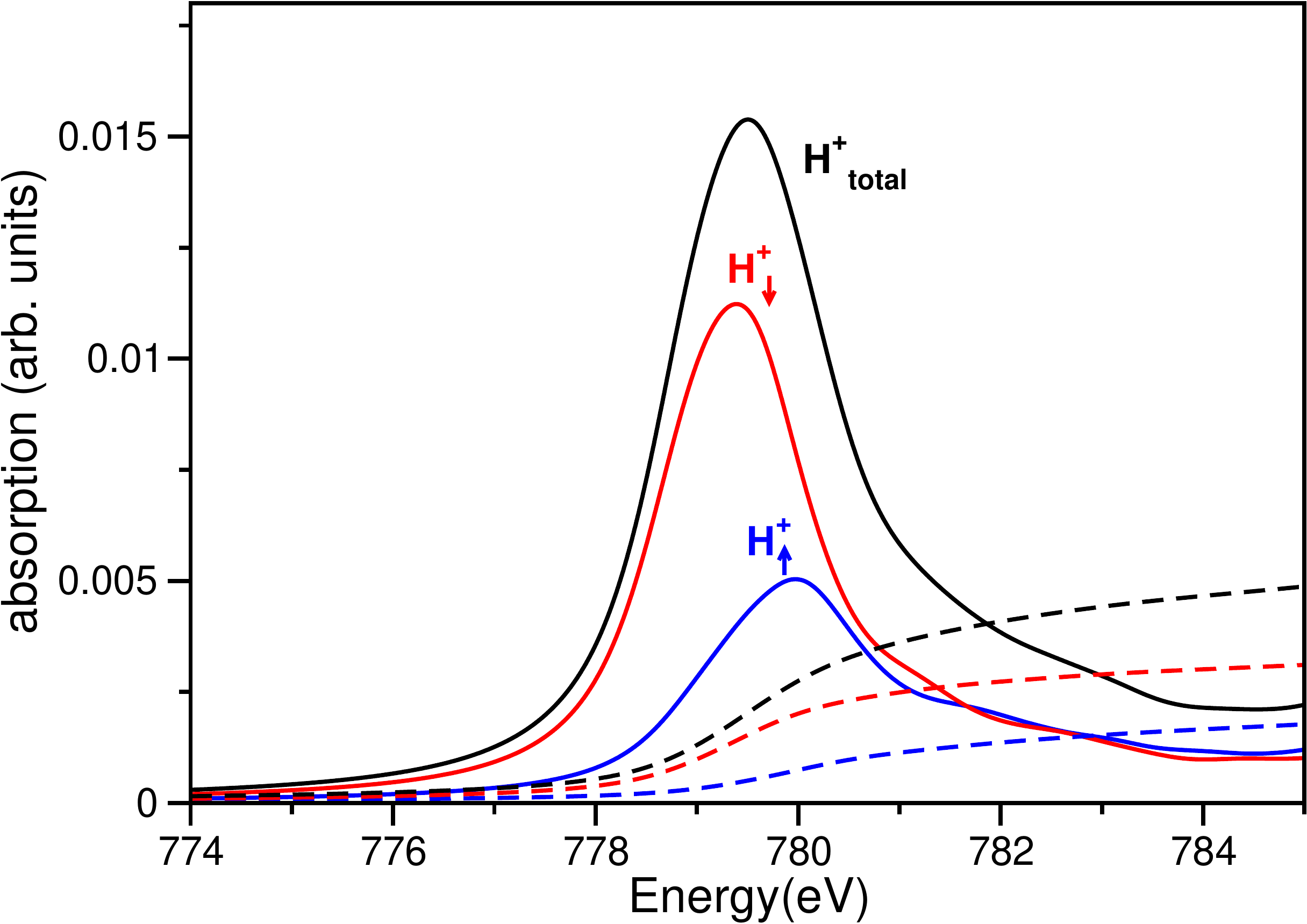}
\caption{Helicity dependent x-ray absorption spectra from right circularly polarized light for Cobalt for the L edge. The response function is calculated in three different ways: by blocking all majority 2$p$ states ($H^+_{\rm min}$), by blocking all minority 2$p$-states ($H^+_{\rm maj}$) and by including all transitions from core levels ($H^+_{\rm total}$, the full response), with the corresponding dashed lines the energy integration of these response functions. The parameters for extracting spin response functions from the response functions of circularly polarized light are given by the ratios of the integrated quantities $\int\! dE\, H^\pm_\sigma(E)/\int\! dE\, H^\pm(E)$ over the L3 edge.
}\label{fig:ab}
\end{figure}

For a realistic solid such a decomposition cannot be rigorously made. However, as we now show \emph{ab-initio} theory allows us to obtain an approximate scheme providing a corresponding decomposition for solids. To this end we consider the spin projected helicity dependent XAS spectra ($H^\pm_\sigma(E,t)$), i.e. the XAS spectra with majority or minority spin transitions blocked, see Fig.~\ref{fig:ab}. We now make the assumption (to be numerically verified later) that an approximation to the average value over a band of the ratios of transition probabilities $p^\pm_{i f\sigma}/\left(\sum_\sigma p^\pm_{i f\sigma}\right)$ can be obtained by integration over the L3 edge the ratio $H^{\pm}_\sigma/H^\pm$. With these ratios in hand an approximation to Eq.~\ref{hxas0} can be written as

\begin{eqnarray}
\label{hxas}
A u(E,t)+B d(E,t)&=\frac{H^+(E,t)}{\int\!dE\, H^{+}(E,0)} \\
A' u(E,t)+B' d(E,t)&=\frac{H^-(E,t)}{\int\!dE\, H^{-}(E,0)}, \nonumber
\end{eqnarray}
where we have employed a second assumption in neglecting any time dependence of these ratios (again, to be validated later by numerical calculation). They thus may be obtained from the ground state as: 

\begin{eqnarray}
A & = & \int\! dE\, H^+_\uparrow(E,0)/\int\! dE\, H^+(E,0), \\
B & = & \int\! dE\, H^+_\downarrow(E,0)/\int\! dE\, H^+(E,0), \nonumber \\
A' & = & \int\! dE\, H^-_\uparrow(E,0)/\int\! dE\, H^-(E,0), \nonumber \\
B' & = & \int\! dE\, H^+_\downarrow(E,0)/\int\! dE\, H^-(E,0) \nonumber. 
\end{eqnarray}
where the energy integration is again over the L3 edge.

In this way we can identify the parts of the response function that come primarily from the bands and their occupation (the $u(E,t)$ and $d(E,t)$) separately from the parts that comes from the transition probabilities ($A$, $B$, $A'$ and $B'$).
From the transient-HXAS response functions of laser pumped systems Eq.~\ref{hxas} can then be solved transiently to obtain $u(E,t)$ and $d(E,t)$, a procedure schematically illustrated in Fig.\ref{fig:scheme}b. This will therefore provide us with the required information: changes in majority (u) and minority (d) charge as a function of energy and time. 

In Fig.~\ref{fig:ab} we shown these response functions and their energy integration over the L3 edge for Cobalt. The calculated values of these parameters of Co found from these response functions are $A=0.846$, $B=0.154$, $A'=0.646$ and $B'=0.355$. For the corresponding values for Fe and Ni see \onlinecite{abc}. In the following sections we now describe the theoretical and experimental method used to determine the dynamics of magnetization from the HXAS spectra. 

\section{Theoretical methodology}

To calculate transient HXAS for laser pumped Co we have used the fully {\it ab-initio} state-of-the-art time dependent density functional theory (TDDFT)\cite{RG1984}, that rigorously maps the computationally intractable problem of interacting electrons to the Kohn-Sham (KS) system of non-interacting fermions in a fictitious potential, which can be solved by modern computing clusters. The time-dependent KS equation is:

\begin{align}
i &\frac{\partial \psi_{j\bk}({\bf r},t)}{\partial t} =
\Bigg[
\frac{1}{2}\left(-i{\nabla} +\frac{1}{c}{\bf A}_{\rm ext}(t)\right)^2 +v_{s}({\bf r},t) \nonumber \\
& + \frac{1}{2c} {\sigma}\cdot{\bf B}_{s}({\bf r},t) + \frac{1}{4c^2} {\sigma}\cdot ({\nabla}v_{s}({\bf r},t) \times -i{\nabla})\Bigg]
\psi_{j\bk}({\bf r},t)
\label{TDKS}
\end{align}
where ${\bf A}_{\rm ext}(t)$ is a vector potential representing the applied laser field. It is assumed that the wavelength of the applied laser is much greater than the size of a unit cell and the dipole approximation can be used, i.e. the spatial dependence of the vector potential is disregarded. The KS potential $v_{s}({\bf r},t) = v_{\rm ext}({\bf r},t)+v_{\rm H}({\bf r},t)+v_{\rm xc}({\bf r},t)$ is decomposed into the external potential $v_{\rm ext}$, the classical electrostatic Hartree potential $v_{\rm H}$ and the exchange-correlation (XC) potential $v_{\rm xc}$. Similarly, the KS magnetic field is written as ${\bf B}_{s}({\bf r},t)={\bf B}_{\rm ext}(t)+{\bf B}_{\rm xc}({\bf r},t)$ where ${\bf B}_{\rm ext}(t)$ is an external magnetic field and ${\bf B}_{\rm xc}({\bf r},t)$ is the exchange-correlation (XC) magnetic field. ${\sigma}$ are the Pauli matrices. The final term of Eq.~\eqref{TDKS} is the spin-orbit coupling term. In the fully non-collinear spin-dependent version of this theory\cite{krieger2015,dewhurst2016} the orbitals, $\psi$ are two component Pauli spinors and using these orbitals magnetisation density can be calculated as:

\begin{eqnarray}\label{st}
{\bf m}({\bf r},t)=\sum_j \psi^*_j({\bf r},t){\m \sigma}\psi_j({\bf r},t),
\end{eqnarray}
the integral of this vector field over space gives the magnetic moment as a function of time (M(t)).

In order to calculate the response function at various times, linear response formalism of the TD-DFT is the used\cite{RG1984,my-book,sharma2011}. This method of calculating the transient response function has been studied extensively\cite{clemens2020,dewhurst2020,sharma22}, and demonstrated to be in very good agreement with experiment. For the response function calculations a broadening of 0.9~eV was used. This choice of smearing is based on ground-state $GW$ calculations, being the average of the width of p$_{1/2}$ and p$_{3/2}$ states. Ground-state $GW$ calculations were also used to determine the position of the 2$p$-states and the KS states were then scissors corrected. These scissors corrected states are used for performing response function calculations. For performing these $GW$ calculations a finite temperature (a temperature of 500~K was used), all electron, spin-polarized $GW$ method\cite{gw} is used, where the spectral function on the real axis is constructed using a Pade approximation. Spin-orbit coupling was included in the $GW$ calculations and a Matsubara cutoff of 12~Ha was used.

In the present work we have used adiabatic local density approximation for XC potential and all calculations are performed using the highly accurate full potential linearized augmented-plane-wave method\cite{singh}, as implemented in the ELK\cite{elk,dewhurst2016} code. A face centred cubic unit cell with lattice parameter of $3.53\si{\angstrom}$ was used for Co. The Brillouin zone was sampled with a $20\times 20\times 20$ k-point mesh. For time propagation the algorithm detailed in Ref.~\onlinecite{dewhurst2016} was used with a time-step of $2.42$ atto-seconds.  The final magnetization value converges with the above mentioned computational parameters to 1.61~$\mu_{\rm B}$. The theoretical pump pulse used in the present work has central frequency of 1.55~eV, duration of 24~fs and incident fluence of 6.7~mJ/cm$^2$. In comparison to the experimental pulse this is of shorter duration, a situation motivated by several factors: (a) non-inclusion of nuclear dynamics prevents us from describing physics at longer time scales and (b) while the method proposed above is state-of-the-art, there is a price to pay in computational burden that, with present computer power, precludes the calculation of longer pulses.    

\section{Experimental method} 

The ultrafast HXAS experiments were conducted at the BESSY~II Femtoslicing facility (beamline UE56/1-ZPM and DynaMaX end-station) at Helmholtz-Zentrum Berlin \cite{Holldack2014}. Femtosecond X-ray pulses of 100~fs temporal width and elliptical polarization are generated via femto-slicing from electron bunches stored in the BESSY~II electron storage ring by interaction with intense infrared laser pulses. Laser pulses from a second laser amplifier system (pump-laser) intrinsically synchronized with the fs-X-ray pulses are used as pump-pulses for sample excitation (wavelength 800~nm and pulse width 50~fs). The effective temporal resolution of this experimental setup is 120~fs. The experiment is operated at a repetition rate of 6~kHz, where only every second fs-X-ray pulse is synchronized to a pump-pulse (i.e. pump-laser running at 3~kHz). This allows for alternatingly probing the excited sample and the sample ground state on a shot-to-shot basis, providing excellent normalization opportunities. X-ray and pump-laser beams enclose an angle of about 1$^{\circ}$ when incident on the sample.
The sample was a Pt(3~nm)/Co(15~nm)/Al(300~nm) layer stack deposited on a 200~nm thin Si$_{3}$N$_{4}$ membrane by magnetron sputtering, which allowed measurement of XAS spectra across the Co L$_{3}$ resonance in transmission geometry. The X-ray photons hit the sample at 43$^{\circ}$ incidence angle to surface normal. The transmitted intensity is detected by an avalanche photo diode screened from the laser light by an Al membrane \cite{Holldack2014}. The relative orientation between X-ray helicity and magnetization was swapped by inversion of an external magnetic field (0.1~T field amplitude) pointing parallel to the X-ray beam. The X-ray absorption spectra were recorded quasi-simultaneously by field inversion for every single photon energy on a ten second basis. The energy resolution of the UE56/1-ZPM beamline achieved at the Co L$_{3}$ absorption edge is approximately 1.6~eV.

To determine time zero of the pump-probe experiment for the recorded transient X-ray spectra with highest possible precision, a MoSi multi-layer reference sample was used\cite{Schick2016,Jana2021}. To correct for slow drifts of time-zero we employed the following measurement procedure: Time zero was determined on the MoSi multi-layer by repeated delay scans with a net acquisition time of 1 hour. The data statistics of this measurement allows determination of time zero in an interval smaller than $\pm$15~fs. Subsequently, X-ray spectra for a set of delays in a $\pm$150~fs interval around nominal time zero from the Co sample were recorded during a 2.5~hours acquisition time. These two types of measurements were alternatingly repeated. The effective delay for any recorded X-ray spectrum was finally recalculated by linear interpolation between the two time-zero values from the measurements on the MoSi multi-layer before and after the XAS acquisitions. Such obtained distribution of effective delays and associated X-ray spectra were finally binned in effective delay windows of 50~fs width ranging from -150 to 200~fs delay, averaged, and evaluated.

\section{Results}

\begin{figure}[ht]
\includegraphics[width=0.9\columnwidth, clip]{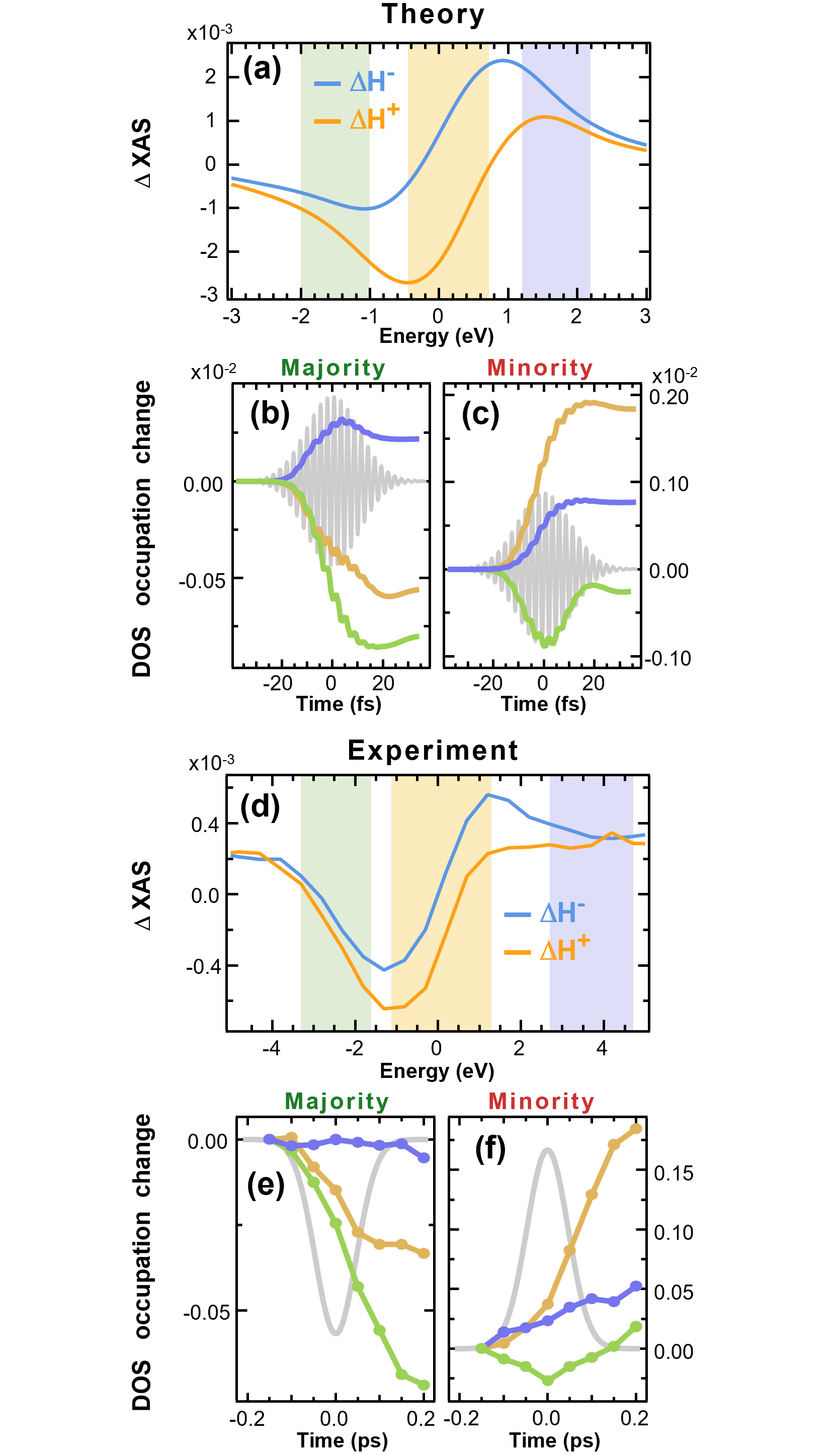}
\caption{Difference between the helicity dependent x-ray absorption spectra (HXAS) before pumping and after laser pumping showing (a) theoretical data and (d) experimental data. The theoretically calculated transient change in the (b) majority and (c) minority spins is obtained by using Eqs.~\ref{hxas} and integrating within the various energy windows, as indicated by different colours in panels shown in (a). For the experiment the integration is performed in the windows shown in (d), and the correspondingly obtained transient majority and minority spins shown in panels (e) and (f) respectively. The pump laser pulse of central wave length 800~nm is shown in grey for theory in panels (b) and (c), while the Gaussian envelope of the experimental pulse is shown in (e) and (f) in grey. Note that the theoretical calculations were performed using a pump pulse of 24~fs duration while in the experiment the laser pulse full width at half maximum is approximately 50~fs. Despite this one can see excellent qualitative agreement between theory and experiment. 
In particular the turning point of the minority charge, that represents the point at which spin-flip induced transitions dominate over spin preserving optical transitions, is seen for both theory and experiment to occur close to half cycle of the pump pulse.
}\label{fig:hxas}
\end{figure}

Employing the method proposed above we now use Co as an example to examine the dynamics of spin and charge upon laser pumping. A pump laser pulse drives demagnetization of Co, a process reflected in the HXAS spectra. These changes are shown in Fig.~\ref{fig:hxas}(a) and \ref{fig:hxas}(d) where we show the difference between the unpumped and pumped HXAS spectra obtained by theoretical calculations (20~fs after laser pumping) and from experiment (50~fs after laser pumping)
respectively. Negative values indicate an increase and positive numbers a decrease in the HXAS signal after laser induced changes in the state occupation (note that a decrease in absorption is equivalent to an increase in state-occupation). As expected, both experiments and theory show a decrease in the state-occupation below and an increase in the state-occupation above Fermi energy (which is set to 0~eV in Fig.~\ref{fig:hxas}(a) and (d)), representing the excitation of charge from below to above the Fermi energy due to the laser pump. The amount of change varies with x-ray helicity, signifying spin-dependent pump effects.

In order to examine changes of majority and minority charge as a function of time and energy, we now solve Eqs.~\ref{hxas} to obtain $u$ and $d$ and integrate these changes of the spin dependent occupation in the energy windows indicated by the shaded rectangles, shown in panels (a) and (d) of Fig.~\ref{fig:hxas}. 
These windows (width and position) are chosen to capture the dynamics of charge loss below the Fermi energy, the dynamics of charge gain above the Fermi energy, and the interesting region around the Fermi energy within which more complex behavior can take place.

\begin{figure}[ht]
\includegraphics[width=\columnwidth, clip]{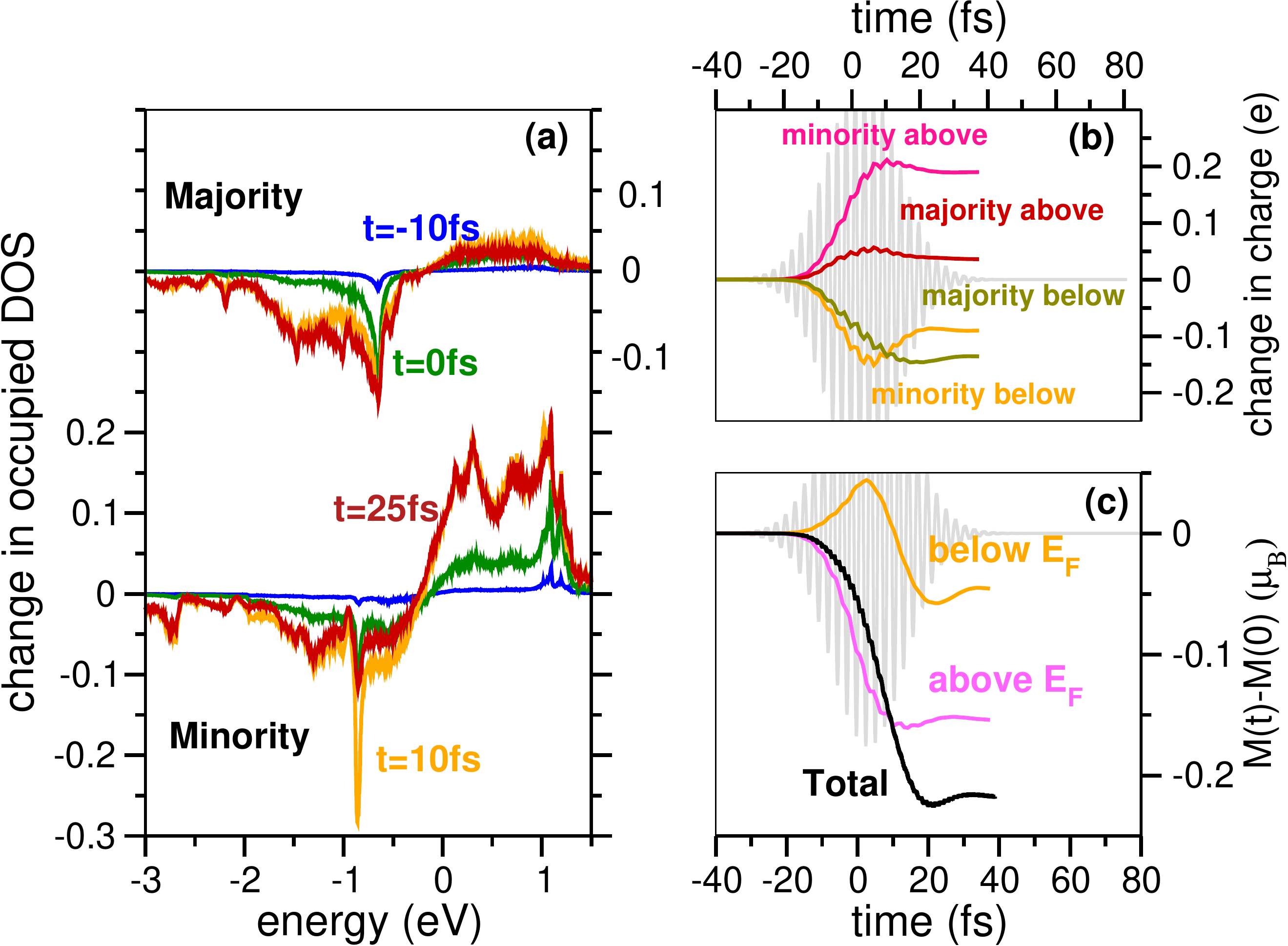}
\caption{(a) Change in transient occupied spin projected DOS. Results are shown at various times after laser pumping. Pump pulse induced (b) change in majority and minority charge integrated below and above the ground-state Fermi energy and (c) change in total spin magnetic moment and spin moment integrated below and above the Fermi energy. The pump pulse has central wave length of 800~nm, full width half maximum of 24~fs and incidence fluence of 6.7mJ/cm~$^2$.
}\label{fig:dos}
\end{figure}

Remarkably, we see that the integrated change in $u$ and $d$ (i.e state occupation) obtained from theory very closely follows changes in the corresponding experimental data(Fig.~\ref{fig:hxas} b, c, e, f), analyzed using exactly the same parameters $A$, $B$, $A'$, $B'$ as derived from and used in analyzing the theoretical data.
Below (green) the Fermi energy both majority and minority state occupations are lost and gained respectively, while for the window enclosing the Fermi energy (yellow) majority is lost and minority gained.
This indicates not just optical excitation of charge but also a large number of spin flips (also see Fig.~\ref{fig:dos} for further details) in this energy range, leading to an increase in the minority occupation and decrease in the majority spins. This leads to demagnetization of Co.  
In the window above the Fermi energy (blue) a strong gain in minority charge carriers is seen both in theory and experiment. In our early time theoretical calculations a much smaller gain in majority is seen, reflecting the nearly full occupation of majority states in the high energy window above $E_F$, while for the longer times used in experiment almost no change in majority charge above $E_F$ is seen. 
We note that both in experiment and theory we find charge excitations outside the energy window of $\pm$1.55~eV around the Fermi energy indicating a contribution from non-linear two-photon excitation processes. To prove this in experiment would entail a full probe of fluence and energy dependent spectra.
 
An interesting feature offering microscopic insight into the spin-flip process is the turning point in the minority charge of the low energy window (green). This turning point precisely demarks the time at which spin-flip processes (which will increase minority charge) begin to dominate over optical excitations (that reduce minority charge) in the low energy window. These results indicate that the spin-flip process play a very important role in the energy region of \emph{low-energy} states, as is indicated by the Hamiltonian which shows that the SOC term is strongest for such low energy localized states as the $\nabla v_s$ term is largest for these states\cite{krieger2015}. 

Despite two different pump pulse lengths used in theory and experiment we see that the qualitative behavior of the spin dependent charge corresponds overall very well between theory and experiment in each energy window. This highlights the universality of the underlying physics of laser pumped demagnetizaton, as well as the robustness of the procedure used to extract spin- and energy- resolved excited charge proposed in the present work.

We note that the energy corresponding to the Fermi level in the experimental data in Fig.~\ref{fig:hxas} is not given in a natural way. We define zero energy in the latter case in analogy to theory at the zero crossing point of the transient $u$ spectrum (see Fig.~\ref{fig:hxas}). The energy resolution of this procedure may be inaccurate by about 0.2~eV; this, however, does not effect any of our conclusions given that the width of our energy window is much broader than this.

The HXAS thus encodes a great deal of information of the underlying excitation processes in ultrafast spin dynamics, well beyond that which can be obtained directly from MCD spectra, which provides only the dynamics of the local moments. While attempts have been made in previous works to infer signatures of such excitation processes from MCD/MLD\cite{LeGuyader2022,clemens2020} spectra, it is only by systematically disentangling the spin degrees of freedom in HXAS, as we do here, that this information can be obtained. To do this the only extra ingredient required are the parameters $A$, $B$, $A'$ and $B'$ in Eqs.~\ref{hxas}. 
As these can be obtained \emph{ab-initio} from the ground state, then this procedure can be preformed for all elemental and multi-component magnets of interest (values of these parameters for Fe, Co and Ni are given in this work).

Having experimental access to energy and spin resolved information via HXAS offers new opportunities to probe in-depth the physics of the excitation process, in particular energy and time resolved changes in magnetization. To explore this physics further in Fig.~\ref{fig:dos}(a) we show changes in the state-density for majority and minority electrons as a function of time. One observes a broad loss of majority states from below the Fermi energy, and a corresponding large gain over and energy range of $\sim1.5$~eV above the Fermi energy. Closer examination unveils a subtle interplay of spin orbit and optically induced transitions, in particular the large peak in the minority density of states difference observed 10~fs but not at 25~fs due to filling of these states by spin-orbit induced transitions from majority to minority. It is this transient peak that is responsible for the turning point in the integrated transient spin resolved DOS occupations seen in both theory and experiment above, and revealed also in the integrated minority charge below Fermi energy, see Fig.~\ref{fig:dos}b. These energy resolved spin channels below and above the Fermi energy also show clear evidence of the extent of spin flips as the increase in minority charge above the Fermi energy is significantly greater than the decrease in minority charge from below the Fermi energy. 

In a similar way we can also energy resolve the dynamics of magnetic moments, 
At early times minority electrons undergo optical excitation from below the Fermi energy, and the magnetic moment below the Fermi energy (orange curve in Fig.~\ref{fig:dos}c) shows a corresponding \emph{increase}. At $\sim20$~fs spin-flip processes begin to dominate, driving a reduction in the majority carriers and hence a reduction in the moment below the Fermi level. At the same time, this excitation of minority charge from below to above the Fermi level leads to a negative change in the moment to develop above the Fermi level (pink). However, the magnitude of the moment above the Fermi energy also starts to saturate as spin-flips dominate the physics of demagnetization. These results again point towards the physics seen in Fig.~\ref{fig:scheme}c and \ref{fig:hxas}; the spin flip process are not just around the Fermi level but are strongly energy dependent and show signatures well below the Fermi energy.

\begin{figure}[ht]
\includegraphics[width=\columnwidth, clip]{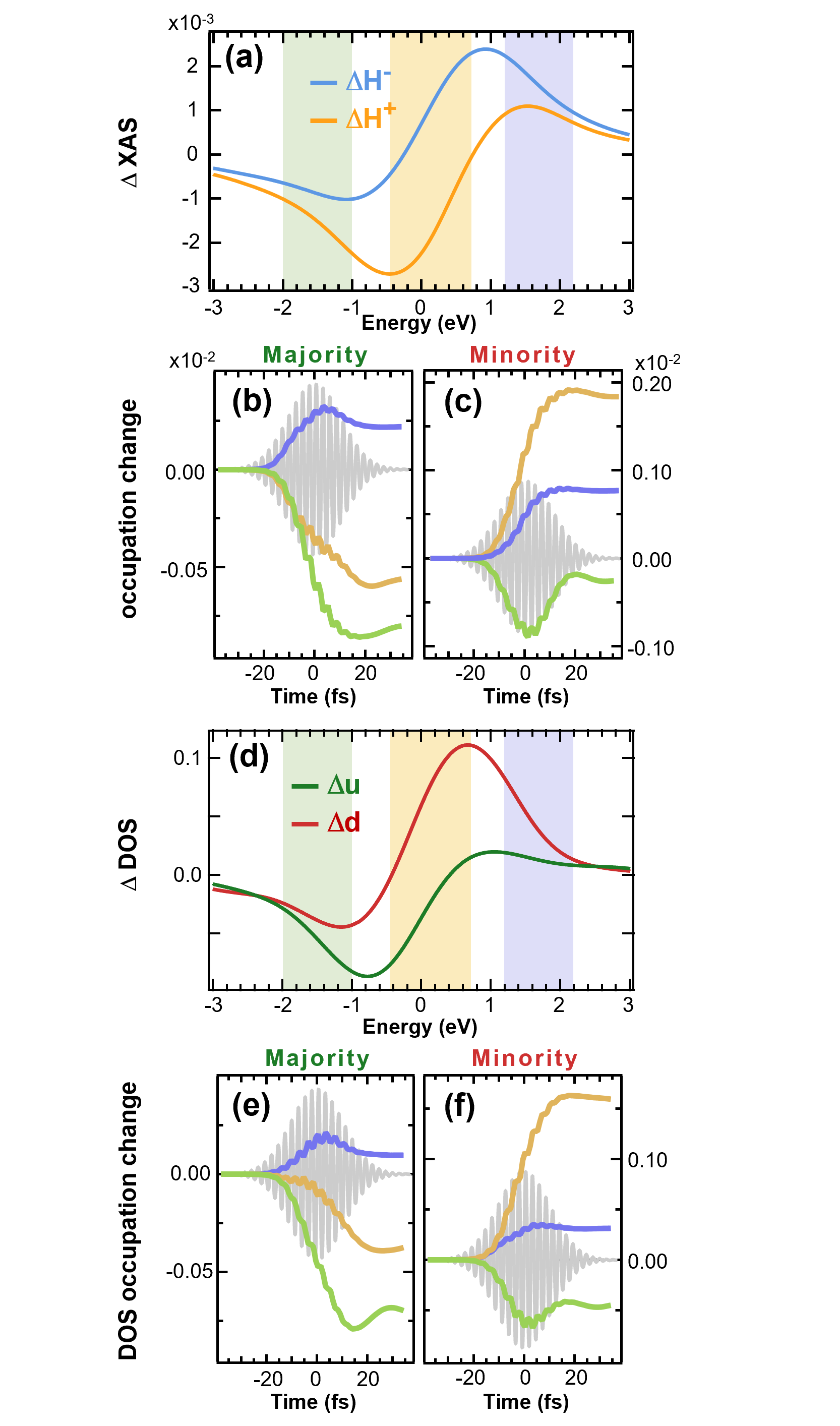}
\caption{Internal theoretical consistency of HXAS and DOS derived spin, energy, and time resolved state occupation. (a) HXAS change 20~fs after laser pumping and (d) the difference in the occupied density of states 20~fs after pumping. Charge in (b) majority and (c) minority spins obtained by using Eqs.~\ref{hxas} and integration in various energy windows indicated by the different colours in panel (a). Change in (e) majority and (f) minority spins obtained from integrating the difference in DOS occupations in the corresponding energy windows shown in panel (d). Note that the energy windows for integration are exactly the same width and at the same position with respect to the Fermi energy in panels (a) and (d). As can be seen, both routes to deriving transient state occupations yield very similar results, validating the method proposed for disentangling spin, energy, and time resolved state occupation from the HXAS spectra. The {\bf A}-field of the pump pulse has an incidence fluence of 6.7mJ/cm~$^2$. 
}\label{fig:th}
\end{figure}

\section{Theoretical validation of HXAS disentanglement}

One of the major approximations in solving Eqs.~\ref{hxas} was the assumption of very weak time dependence of the transition probabilities. This not only reduces the computational task, most importantly it allows calculation and tabulation of $A, B, A'$ and $B'$ from static ground-state calculations for the analysis of future experiments. While we have justified this assumption {\it post-hoc} by agreement between theory and experiment, within the theoretical framework itself this approximation can be justified. In order to do this in Fig.~\ref{fig:th} we present the changes in HXAS and DOS occupations 20~fs after laser pumping. The DOS is broadened by a Gaussian (FWHM $0.82$~eV) in order to have the same spectral width as the HXAS response function. Evidently, the two sets of data show the same trend in that occupation of both majority and minority spins decreases below and increases above the Fermi energy, while at the Fermi level the two curves cross zero. Closer inspection of the helicity dependent response function, Fig.~\ref{fig:th}(a), and the transient DOS occupation, Fig.~\ref{fig:th}(d), reveals slight differences arising from the fact that the response functions contain information both of changes in occupation as well as the respective transition probabilities.
Integrated changes in the occupation in three differently positioned energy windows are shown in panels \ref{fig:th}(b) and (e) for majority and (c) and (f) for minority spins. In this case, we see that the integrated charge from HXAS very closely follows changes in the corresponding state density. In both the positive (blue) and negative (green) energy windows both majority and minority states are lost and gained respectively, while for the window enclosing the Fermi energy (yellow) majority is lost and minority gained. This close agreement between the changes in spin projected occupations obtained from DOS and HXAS calculations provides a demonstration within the theoretical framework that the extraction from the HXAS of pure spin response functions using coefficients tabulated from the ground state is indeed a good approximation. 

\section{Discussion}

Strong light-matter interaction in magnetic systems creates a highly non-equilibrium state of matter exhibiting remarkable phenomena such as all-optical switching and OISTR. Exploring this diversity of phenomena demands experimental probes able to precisely track the microscopic processes underlying ultrafast spin dynamics. In the present work we have proposed and verified a simple extension of the well established XMCD technique, allowing a transient spin- energy- and species-resolved probe of ultrafast demagnetization.

This extension is based upon disentangling the spin degrees of freedom that are, generally, woven in a complex way into the response of a magnetic system to circularly polarized light. We show that, by employing parameters obtained \emph{ab-initio}, the experimental HXAS spectra may be inverted to give spin specific response functions that, in turn, provide a window into the spin-energy landscape in which ultrafast demagnetization occurs.

Using the example of elemental cobalt we resolve the magnetization dynamics into energy regions that vividly illustrate the wide energy range on which the spin-orbit and optical transitions that drive demagnetization occur. Specific quantitative features, such as turning points in spin and energy resolved charge, further reveal critical points at which the balance between spin preserving optical transitions and spin-orbit induced spin-flips changes.

The \emph{ab-initio} procedure employed to disentangle the HXAS spectra is shown to be valid within the theoretically obtained HXAS spectra and state density, with the qualitative agreement between theory and experiment (which employ lasers pulses of significantly longer duration than those that can be treated theoretically) further bolstering the correctness of the spin projection of the HXAS spectra. As the parameters encoding spin specific transitions required to preform this transformation are, to an excellent approximation, time independent, they can be obtained from the ground state and so this procedure can, in principle, be applied to multi-component magnetic materials of almost any complexity. The extension of HXAS that we propose is therefore of great applicability.

Promising future experimental applications of this "extended HXAS" method include probing spin- and energy-resolved dynamics in all-optical spin switching magnets. Direct access to the transient spin- and energy resolved charge is expected to provide invaluable insight into still controversial features of the dynamics of spin switching, e.g. which moments rotate, at what times, and the in-plane versus out-of-plane components of magnetization. Understanding the still unclear microscopic world underpinning such well established phenomena may provide the basis for enriched exploration and control of the frontiers of ultrafast spin dynamics.

\section{Acknowledgements}

Sharma, CvKS, SE, NP and CSL would like to thank for funding by the Deutsche Forschungsgemeinschaft (DFG, German Research Foundation) – Project-ID 328545488 – TRR 227 (projects  A04, A02 and A03).  
Shallcross would like to thank DFG for funding through SPP 1840 QUTIF Grant No. SH 498/3-1. The authors acknowledge the North-German Supercomputing Alliance (HLRN) for providing HPC resources that have contributed to the research results reported in this paper.

%

\end{document}